\documentclass[article]{aa}  
\usepackage{graphicx}
\usepackage{txfonts}
\usepackage{natbib}
\bibpunct{(}{)}{;}{a}{}{,} % to follow the A&A style

\begin{document}

\titlerunning{Cloud reflection model for Jovian impact flashes}
\authorrunning{Arimatsu et al.}

   \title{Cloud reflection modelling for impact flashes on Jupiter}
    \subtitle{A new constraint on the bulk properties of the impact objects}

   \author{Ko Arimatsu\inst{1}
          \and
          Kohji Tsumura\inst{2}
          \and
          Fumihiko Usui\inst{3}
          \and
          Jun-ichi Watanabe\inst{4}
          }

   \institute{The Hakubi Center / Astronomical Observatory, Graduate School of Science,  Kyoto University Kitashirakawa-oiwake-cho, Sakyo-ku, Kyoto 606-8502, Japan\\
              \email{arimatsu.ko.6x@kyoto-u.ac.jp}
                \and
                Department of Natural Science, Faculty of Science and Engineering, Tokyo City University, Setagaya, Tokyo 158-8557, Japan
                \and
                Institute of Space and Astronautical Science (ISAS), Japan Aerospace Exploration Agency (JAXA), 3-1-1 Yoshinodai, Chuo-ku, Sagamihara, Kanagawa 252-5210, Japan
                \and
                Astronomy Data Center, National Astronomical Observatory of Japan 2-21-1 Osawa, Mitaka, Tokyo 181-8588, Japan
             }

    \date{}

\abstract
%{Modelling of the cloud reflections provides accurate bulk properties of impact flash events on Jupiter.}
{}
{We investigate optical characteristics of flashes caused by impacting meter- to decameter-sized outer solar system objects on Jupiter and contributions of reflected light from surface clouds at visible wavelengths to estimate more accurate bulk parameters such as the luminous energy of the flash, the kinetic energy, the mass, and the size of the impact object.}
{Based on the results of recent reflectivity studies of the Jovian surface, we develop a cloud reflection model that calculates the contribution of the reflected light relative to that directly from the flash. 
We compare the apparent luminous energy of the previously reported flashes with the expected cloud reflection contributions to obtain their revised bulk parameters. 
}
{We found that the cloud reflection contributions can be up to 200\% of the flux directly from the flash %, 
and thus can be the most significant uncertainty in the measurement of the bulk parameters. 
The reflection contributions strongly depend on wavelength.
With our cloud reflection correction, the revised bulk parameters of the previously reported flashes are obtained. }
{
Our cloud reflection correction provides a better understanding of the properties of impacting objects on Jupiter and is crucial for ongoing detailed investigations using high-sensitivity and multi-wavelength observation systems such as PONCOTS. 
It will also be useful for understanding other optical transients in Jupiter's upper atmosphere, such as the recently discovered sprite-like events.}


\maketitle      

\section{Introduction} \label{sec:intro}
After the first discovery by \citet{Hueso2010-xv}, second-timescale optical flashes on Jupiter have been serendipitously detected by ground-based amateur observers.
Previous studies by \citet{Hueso2010-xv, Hueso2013-ki,Hueso2018-nr} and \citet{Sankar2020-lm} carried out photometric analyses of the recorded video data and investigated their emission characteristics.
The optical energy estimates based on the photometric analyses indicate that the observed flashes are due to the impact of unidentified interplanetary objects with sizes in the meter-to-decameter range.
Their energy estimates, therefore, provide a unique opportunity to study the abundance and physical properties of small objects in the outer solar system. 
Furthermore, the emission properties of the Jovian flashes provide the radiation characteristics of large impacts on atmospheres, which could potentially threaten human society \citep{Jenniskens2019-tg,Boslough1997-pi,Boslough2008-ae}, but are still unknown due to their infrequent occurrence on Earth \citep{Brown2002-aq}.

Unlike lunar impact flashes, which occur on the solid surface of the Moon (e.g., \citealt{Ortiz2000-za,Avdellidou2019-ks}), impact flashes of planets with a substantial atmosphere, such as Jupiter, occur in the stratosphere \citep{Sankar2020-lm}.
The Jovian flashes illuminate the surfaces of the Jovian upper clouds at $\sim 600$~mbar, which are approximately $30\--50$ km below the illumination source, and reflections of light from the clouds can contaminate the observed emission \citep{Borovicka2009-hw}.
In the previous impact flash studies by \citet{Hueso2013-ki,Hueso2018-nr}, the contribution of the cloud reflection component was corrected with a constant correction factor.
This "classical" correction method is useful for order-of-magnitude estimates of the bulk properties (energy, mass, size) of the flashes.
However, this method may become insufficient for recent and near-future multispectral and high-precision photometric studies, 
because the real contribution of the cloud reflection can strongly depend on the wavelength and the geometric arrangement of the observations, expressed as the emission angle, which is the angle between the zenith of the Jovian impact site and the observer (e.g., \citealt{Li2018-uj}).

 A recent impact flash on Jupiter was detected by an optical observation system dedicated to the Jovian flashes, Planetary ObservatioN
Camera for Optical Transient Surveys (PONCOTS; \citealt{Arimatsu2022-if, Arimatsu2023-gz}) on 15 October 2021.
As PONCOTS is a multi-band optical high-cadence imaging observation system, 
three-wavelength high-cadence images of the Jovian flash were obtained for the first time.
These three-band images were calibrated with a spectrophotometric
standard star and provide a photometric record of the Jovian flash with unprecedented precision.  
The obtained spectral energy distributions (SEDs) show a strong excess feature in wavelength bands where a significant cloud reflection is expected (\citealt{Arimatsu2022-if}, see also Sect.~\ref{sub:PONCOTS}). 
Detailed analyses of Jovian flashes using multi-wavelength data sets therefore require modelling and investigation of the wavelength and emission angle-dependent cloud contributions.
The recent development of high-cadence multiband instruments dedicated to flash observations, such as PONCOTS, and the improvement in sensitivity of industrial CMOS sensors adopted by amateur observers (e.g., \citealt{Arimatsu2017-lg,Arimatsu2019-zk}) will allow more frequent multiband detections of Jovian flashes.    
The demand for appropriate modelling of cloud reflections is expected to increase even more in the near future.

In this paper, we propose a new cloud reflection model that takes into account the wavelength and emission angle dependence of reflectivity, and investigate the contributions of the cloud reflection component to the observed flash brightnesses. 
Recent studies of global images of Jupiter taken by the Cassini spacecraft have succeeded in producing a precise multi-wavelength phase curve model of the Jovian surface \citep{Li2018-uj, Heng2021-my}. 
These results provide an unprecedented opportunity to understand the characteristics of cloud reflection and to develop an alternative reflection model sufficient for more detailed flash analyses.
We also present the results of applying our proposed cloud reflection model to the previously reported Jovian impact flashes to demonstrate the importance of accurately correcting for the cloud reflection component in impact flash analyses.
Section~\ref{sec:model} presents our proposed cloud reflection model.
The results and discussion of the application of the cloud reflection model to our 2021 flash detections and the other previous studies are presented in Sect.~\ref{sec:results}.
Finally, the results and discussions are summarised in Sect.~\ref{sec:concl}.

\section{Cloud reflection model}
\label{sec:model}

Impact flash radiation reflected from Jupiter's upper clouds can contribute to the observed fluxes
$F_{\mathrm{obs}}$. To obtain corrected fluxes, $F_{\mathrm{cor}}$, we derive a cloud-reflectivity contribution factor, $f_{\mathrm{CR}}$, given by
\begin{equation}
F_{\mathrm{cor}} = \frac{1}{1+ f_{\mathrm{CR}}} \, F_{\mathrm{obs}}.
\end{equation}
$f_{\rm CR}$ is the ratio of the observed flux from the cloud reflection to that directly from the source.
The previous studies by \citet{Hueso2013-ki,Hueso2018-nr} used a constant correction factor for the cloud reflection of $f_{\rm CR} = 0.3$ under simple and intuitively acceptable assumptions; almost 50\% of the light from a flash illuminates the Jovian upper clouds and is reflected with an approximate albedo of $\sim 0.5$.
However, 
the observed reflected light comes from an entire surface area illuminated by the flash and can be stronger than the emission directly from the emission source.

\begin{figure}[ht!]
\centering
\includegraphics[width=\hsize]{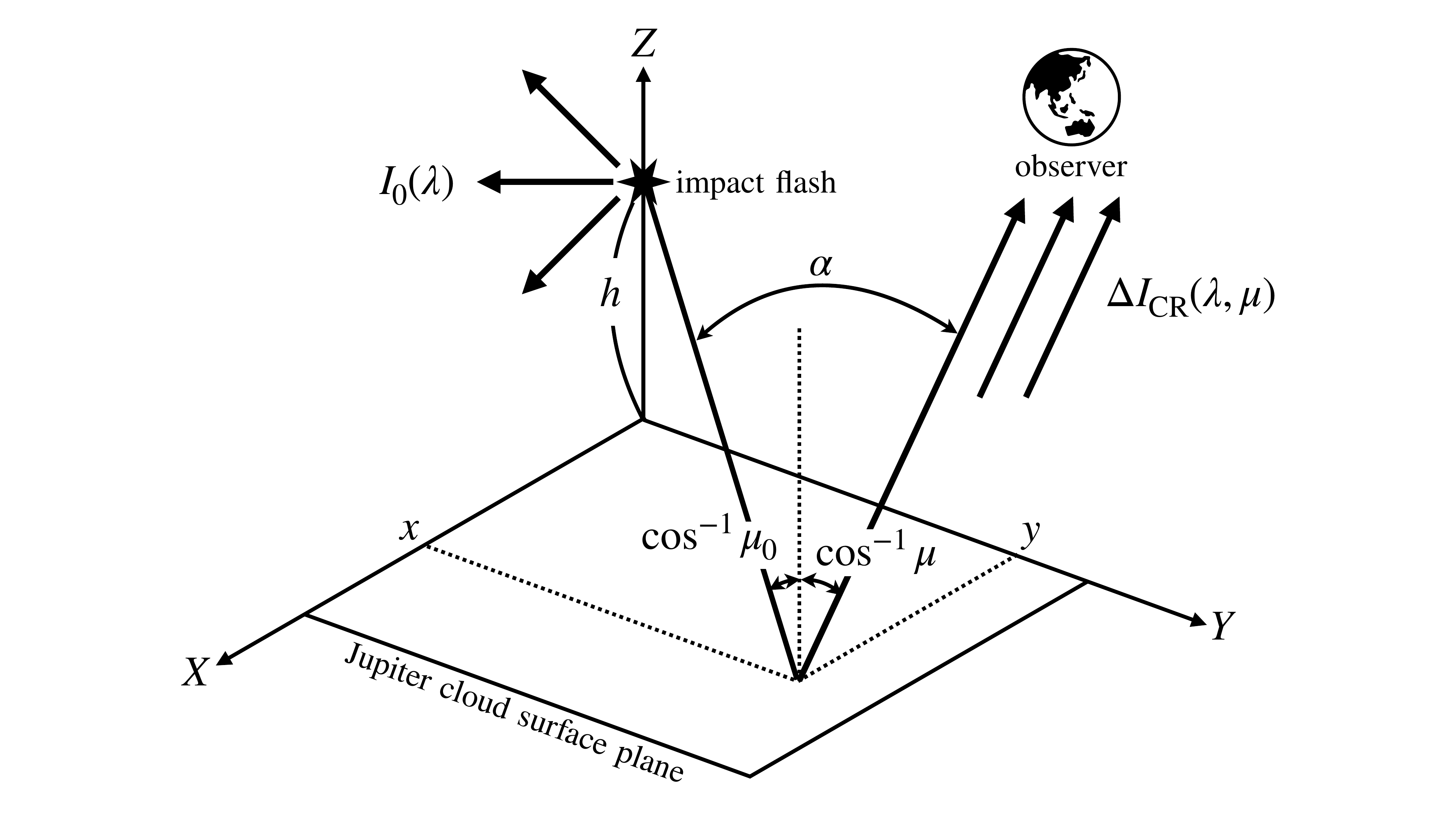}
\caption{Notation used in the present study.
\label{fig1}
}
\end{figure}

In our modelling, we assume that the Jovian clouds reflect light as a Lambertian surface.
We consider a radiation source at height $h$ above flat clouds distributed on a $x\-- y$ plane (Fig.~\ref{fig1}).
With the isotropic source intensity per unit solid angle
$I_0$ and the reflected power radiated $\Delta I_{\rm CR}(\mu)$
at the emission angle $\cos^{-1}{\mu}$, 
$f_{\rm CR}$ is given by $f_{\rm CR} = \Delta I_{\rm CR}(\mu)/I_0$.
The incident light per unit surface area at $(x, y, 0)$ is 
$\mu_0 I_0 /(x^2 + y^2 + h^2)$,
where $\mu_0$ is the cosine of incidence angle, i.e., $\mu_0 = h/\sqrt{x^2 + y^2 +h^2}$.
If we assume the Lambertian reflectance of the Jovian surface with Lambertian albedo $p_L$, 
the cloud reflection component $\Delta I_{\rm CR}(\mu)$ is given with a radial distance from $(0, 0, 0)$, i.e., $r = \sqrt{x^2 + y^2}$, by
\begin{eqnarray}
    \Delta I_{\rm CR}(\mu) & = & \mu  \int^{+\infty}_{-\infty} \int^{+\infty}_{-\infty} \frac{p_L}{\pi} \frac{\mu_0 \,  I_0} {x^2 + y^2 + h^2}  ~{\rm d}x \, {\rm d}y \\
     & = & 2\, \mu \, p_L \, I_0 \int^{+\infty}_{0} \frac{h} {(r^2 + h^2)^{3/2}} \, r ~{\rm d}r \\
     & = & 2 \, \mu \, p_L \, I_0 \\
     & = & 3 \, \mu \, p_g \, I_0, \label{equ_lamb}
\end{eqnarray}
where $p_g$ is a geometric albedo with the relation 
$p_g = (2/3) \, p_L$
for a Lambertian sphere.
$f_{\rm CR}$ is thus expected to be up to $\sim 3$ times of the geometric albedo.
According to recent Jovian reflectivity studies by \citet{Li2018-uj},
$p_g$ reaches its maximum value of $p_g \sim 0.65$ at an optical wavelength range of $\lambda = 500 \-- 700 ~ {\rm nm}$.
$f_{\rm CR}$ can therefore be up to $\sim 3 \times 0.65 \sim 2$ at $\mu \sim 1$.

In reality, the reflectivity of the Jovian surface clouds depends strongly on the reflection angle and the observed wavelength (e.g., \citealt{Li2018-uj}).
The reflected component is maximised for low latitude impacts near the central meridian and drops to zero for those at the limb.
We then derived $\Delta I_{\rm CR}(\lambda, \mu)$ relative to $I_{0}(\lambda)$ based on data sets of the wavelength-dependent Jovian surface reflection phase functions obtained by \citet{Li2018-uj} and \citet{Heng2021-my}.
The $\Delta I_{\rm CR}(\lambda, \mu)$ was estimated by integrating the light rays from the source $I_{0}(\lambda)$ reflected by each surface element of the clouds with incidence ($\cos^{-1}{\mu_0}$), 
emission ($\cos^{-1}{\mu}$), and phase ($\alpha$) angles as follows:
\scriptsize
\begin{equation}
\Delta I_{\rm CR}(\lambda, \mu) = \mu \int^{+\infty}_{-\infty} \int^{+\infty}_{-\infty} \frac{I_{0}(\lambda)}{x^2+y^2+h^2} \frac{\omega}{4} \frac{\mu_0}{\mu_0 + \mu} \big( P(\alpha) - 1 + H(\mu) \, H(\mu_0) \big) ~{\rm d}x \, {\rm d}y,
\end{equation}
\normalsize
where $P(\alpha)$ is the phase function and $H (\mu)$ is the Chandrasekhar $H$-function representing multiple isotropic scattering, which satisfies the following integral equation:
\begin{equation}
H\left(\mu\right)=1+\frac{1}{2}\omega\, \mu\, H\left(\mu\right)\int_{0}^{1}{\frac{H\left(\mu^\prime\right)}{\mu+\mu^\prime} {\rm d}\mu^\prime},
\end{equation}
where $\omega$ is a single-scattering albedo.

\begin{figure}[ht!]
\centering
\includegraphics[width=\hsize]{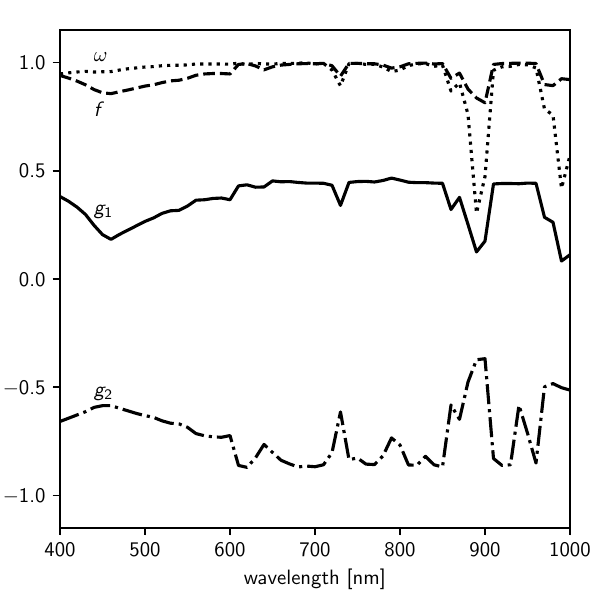}
\caption{DHG parameters used in the present study as a function of wavelength derived by \citet{Heng2021-my}.
Scattering asymmetry factors $g_1$ and $g_2$, weighting factor $f$, and single-scattering albedo $\omega$ are shown as solid, dotted dashed, dashed, and dotted lines, respectively.
\label{fig2}
}
\end{figure}

\begin{figure}[ht!]
\centering
\includegraphics[width=\hsize]{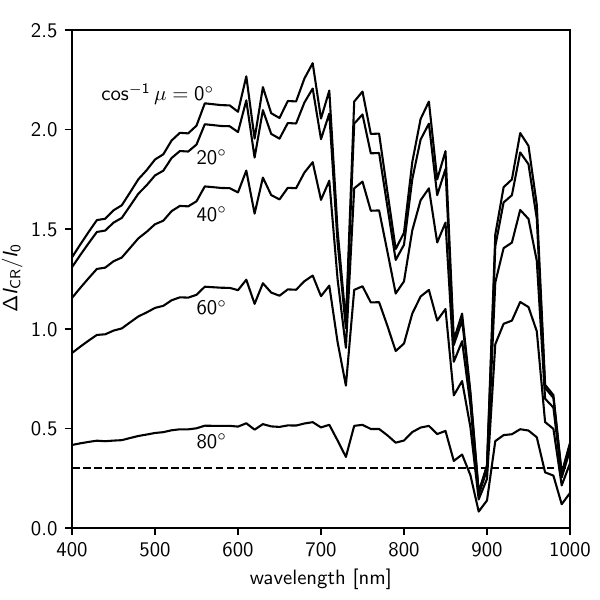}
\caption{$\Delta I_{\rm CR}(\mu)/I_0$ as a function of wavelength for different emission angles $\cos^{-1}{\mu}$.
The horizontal dashed line represents a $f_{\rm CR}$ value used for previous studies ($f_{\rm CR} = 0.3$; \citealt{Hueso2013-ki,Hueso2018-nr}).
\label{fig3}
}
\end{figure}

For the phase function $P(\alpha)$, we use the Double Henyey–Greenstein (DHG) scattering phase functions given by
\begin{equation}
P(\alpha) =\frac{f\ (1-g_1^2)}{(1+g_1^2+2 \, g_1 \cos{\alpha})^{3/2}}+\frac{(1-f)(1- g_2^2)}{(1+g_2^2+2 \, g_2 \cos{\alpha})^{3/2}},
\end{equation}
where $g_1$, $g_2$ are scattering asymmetry factors, and $f$ is a fractional factor for each Henyey–Greenstein function. 
These parameters ($g_1$, $g_2$, $f$ and $\omega$) for the DHG functions have recently been estimated by \citet{Heng2021-my}. 
They used reflected light phase curves of Jupiter derived from the multi-wavelength global images with different phase angles taken by the Cassini spacecraft.
The four DHG parameters as a function of wavelength are shown in Fig.~\ref{fig2}.
The derived $\Delta I_{\rm CR}/I_0$ as a function of $\lambda$ for different emission angles $\cos^{-1}{\mu}$ is shown in Fig.~\ref{fig3}.
For $\cos^{-1}{\mu} = 0$ ($\mu = 1$), $\Delta I_{\rm CR}/I_0$ becomes up to $\sim 2$ at $\lambda \sim 500 \-- 700~{\rm nm}$, 
where $p_g$ reaches a maximum value \citep{Li2018-uj}. 
This maximum is roughly in agreement with that expected from the isotropic reflectivity model 
($f_{\rm CR} \sim 3 \times p_g \sim 2$ at $\mu \sim 1$, see Eq.~\ref{equ_lamb})
but approximately an order of magnitude larger than the previous assumption ($f_{\rm CR} \sim 0.3$; \citealt{Hueso2013-ki,Hueso2018-nr}).
Even for smaller $\mu$ cases ($\cos^{-1}{\mu} < 80\degr$), $\Delta I_{\rm CR}/I_0$ is larger than $0.3$ in most wavelength ranges. 
On the other hand, 
$\Delta I_{\rm CR}/I_0$ is smaller than 0.3 for $\mu = 0 \-- 1$ at $\lambda \sim 890~{\rm nm}$, where a strong methane absorption band is present, and $p_g$ is exceptionally low (e.g., \citealt{Karkoschka1994-qq, Karkoschka2010-vn}).

The present model assumes an infinite illumination area of flat surface and may overestimate the intensity of the reflection component because a real flash of finite height illuminates a finite area within the visible horizon due to the curvature of the Jovian surface.
 It is difficult to make an accurate estimate of the flash height from the observed light curve, which depends strongly on the material properties of the impact object.   
According to the previous simulation results of the observed Jovian flashes 
\citep{Sankar2020-lm,Arimatsu2023-gz}, 
the typical height of the flashes above the cloud surface is estimated to be 20 km or more. 
Assuming a flash height to be 20 km as the lower limit, $\Delta I_{\rm CR}/I_0$ is up to 2 \% smaller than that derived from our original model calculation. This difference can lead to underestimation of $F_{\rm cor}$ of up to 1 \% in the present calculation and is negligible in the present studies.

The present model also ignores the finite size of the flash radiation area, which can have a non-negligible effect on the reflection component calculations.
Based on the typical luminosity $L$ ($L \sim$ a few of $10^{14}$ W or less, \citealt{Hueso2018-nr}) and the effective temperature $T$ ($T = 6500 \-- 10000$ K, \citealt{Hueso2010-xv,Giles2021-sj,Arimatsu2022-if}) of the previously observed Jupiter flashes, 
the total effective area emitting the radiation is approximated to be $L/(\sigma \, T^4)$, 
where $\sigma$ is the Stefan-Boltzmann constant, and derived to be $\lesssim 3\, {\rm km^{-2}}$. 
Assuming the radiation source is spherical-like, 
its typical scale is therefore $\lesssim 1\, {\rm km}$, which is one to two orders of magnitude smaller than the typical height of the flash.
The observed flash can thus be approximated as a point-like source for typical cases.

$f_{\mathrm{CR}}$ for each observation is obtained by integrating $\Delta I_{\rm CR}/I_0$ over wavelengths as follows: 
\begin{equation}
f_{\mathrm{CR}}=  \frac{\int R(\lambda) \, \Delta I_{\rm CR}(\lambda, \mu) \, {\rm d} \lambda}{\int R(\lambda) \, I_{0}(\lambda) \, {\rm d} \lambda},
\end{equation}
where $R(\lambda)$ is the system response of an observation instrument at wavelength $\lambda$.

\section{Results and discussion}
\label{sec:results}

\subsection{Applications of the reflection model to the 2021 October flash event (PONCOTS flash)}
\label{sub:PONCOTS}
\begin{figure}[ht!]%[ht!]
\centering
\includegraphics[height=120mm]{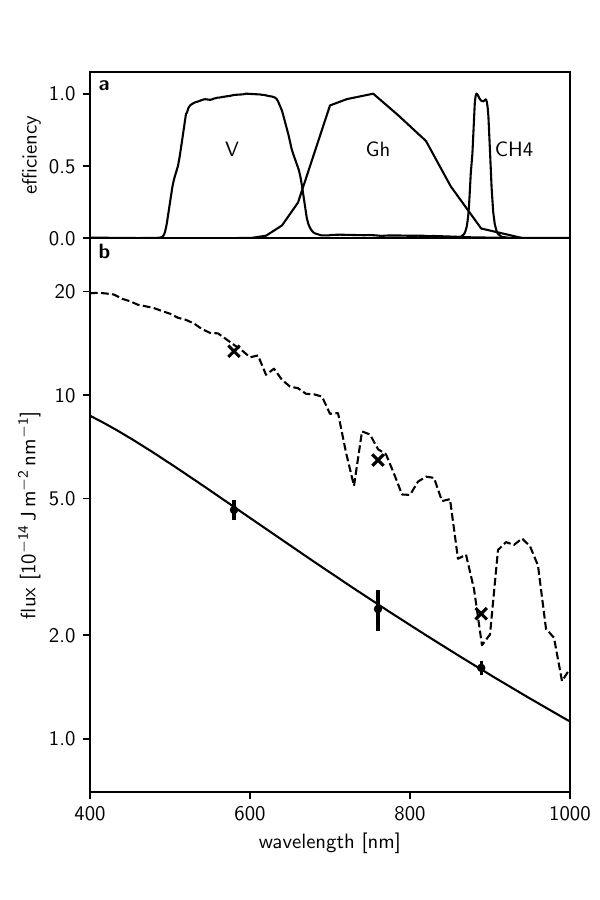}
\caption{Example of the cloud reflection correction results of the 2021 October impact flash (PONCOTS flash) data. (a) Relative spectral responses of the three PONCOTS bands; $V$, Gh, and ${\rm CH_4}$ bands. 
(b) a SED obtained during the peak phase of the flash overlaid with its cloud correction results and the best-fit model spectrum.
Crosses and points with error bars represent observed fluxes and those after cloud-reflection correction, respectively. 
The best-fit spectrum of single-temperature blackbody radiation with a best fit temperature of 8300~K is shown as the solid line. 
Dashed line corresponds to the expected total observed flash spectrum including the cloud reflection contribution.
\label{fig4}
}
\end{figure}

On October 15, 2021, a bright optical flash was detected with the PONCOTS observation system \citep{Arimatsu2022-if}, 
which is dedicated to observing Jovian impact flashes.
PONCOTS observed the flash simultaneously in three different wavebands, 
the V, Gh, and ${\rm CH_4}$ bands with effective wavelengths of $505 \-- 650$, $680 \-- 840$, and $880 \-- 900$ nm, respectively (Fig.~\ref{fig4}a).
As mentioned in \citet{Arimatsu2022-if}, the three band fluxes obtained with the PONCOTS observation system were calibrated using the present cloud reflection model.

Figure~\ref{fig4}b shows an example of the application of the cloud reflection correction to the observed PONCOTS flash SEDs.
The observed fluxes (crosses in Fig.~\ref{fig4}b) show a strong excess in the $V$ and Gh bands relative to the ${\rm CH_4}$ band.
Without the spectral-dependent correction for cloud reflections, 
single or multiple-temperature blackbody spectral models cannot be fitted to the observed SED because its spectral slope is steeper than that of Rayleigh-Jeans.  
With our cloud reflection model, $f_{\rm CR}$ at $\mu = 0.91$ (the emission angle for the PONCOTS flash) for the PONCOTS V, Gh, and ${\rm CH_4}$ bands are estimated to be 1.9, 1.7, and 0.4, respectively.
After subtracting the cloud reflection contributions, the apparent excess in the SED disappears (points with error bars in Fig.~\ref{fig4}).
A single-temperature blackbody radiation spectral model approximates the corrected SED well with a best fit temperature of $\sim 8300$ K \citep{Arimatsu2022-if}.
The total luminous energy $E_0$ is obtained from the SED fit results to be $E_0 = 18^{+9}_{-2} \times 10^{14} ~ {\rm J}$.
The total kinetic energy $E_T$ was derived from $E_0$ using the relationship taken from \citet{Brown2002-aq},
\begin{eqnarray}
E_T & = & \eta^{-1} \, E_0 \\
\eta & = & 0.12 \, E_0^{0.115},
\end{eqnarray}
where $\eta$ is the optical luminous energy efficiency, and $E_T$ and $E_0$ are in kilotons of TNT (kt; $1\, {\rm kt} = 4.19 \times 10^{12} \,  {\rm J}$).
For the PONCOTS flash, $E_T$ is derived to be $E_T = 74^{+33}_{-9} \times 10^{14} ~ {\rm J}$.
Since the impact velocity $v_0$ for the impact object is assumed to be comparable with the escape velocity of Jupiter \citep{Harrington2004-kf}, i.e., $v_0 \simeq 60~{\rm km \, s^{-1}}$, 
the mass of the impact object $M_0$ was given with $E_T$ by 
$M = 2~E_T / v_0^2= 4.1^{+1.9}_{-0.5} \times 10^6 \, {\rm kg}$. 
Assuming its spherical shape, the diameter of the object $D$ is described with a volume density $\rho$ to be $D = (6\, M/(\pi \, \rho))^{1/3}$.
In this study, $\rho$ is assumed to be $\rho = 600 ~ {\rm kg \, m^{-3}}$ for reference, and $D$ is then derived to be $D = 23.5^{+3.1}_{-1.0} ~ {\rm m}$.
The obtained bulk parameters are tabulated in Table~\ref{tab:Analysis}.

\subsection{Applications of the reflection model to the previously reported flash events}
As of April 2023, 7 flashes, including the PONCOTS flash, have been detected by ground-based optical instruments and reported in previous studies.
This subsection provides a brief review of the previously reported flashes and demonstrates the importance of our developed cloud reflection model for the bulk properties of their impact objects. 
As noted in Sect.~\ref{sec:intro}, the contribution of the cloud reflection component has been corrected with a constant correction factor in the previous flash studies.
It should be noted that the previous approximation provides sufficient accuracy for order of magnitude estimates of the bulk properties of the impact objects.
On the other hand, applications of our reflection model to the previous flash events should provide opportunities to demonstrate how the cloud reflection contribution dominates the observed flash brightness and is of great importance for more detailed studies of impact objects.

Table~\ref{tab:Analysis} summarises the results of the re-analysis of the previously reported flashes.
Our revised $f_{\rm CR}$ values are significantly larger than that previously assumed ($f_{\rm CR} \sim 0.3$),
and corrected the $E_0$, $E_T$, $M$ values of the flashes by up to a factor of two.
Since $D$ is proportional to $M^{1/3}$, 
our cloud reflection correction does not significantly change the impact size estimation results (up to $\sim 30\%$ of the original values)
or the conclusions of the previous impact size frequency studies \citep{Hueso2013-ki,Hueso2018-nr}.
On the other hand, this size correction implies that the frequency of very large impacts leaving observable debris fields in the Jovian atmosphere would be lower than previously thought ($0.4 \-- 2.6 ~{\rm events ~yr^{-1}}$; \citealt{Hueso2018-nr}).
In fact, the PONCOTS flash is estimated to be the largest impact observed since 2010, and it was not large enough to leave a debris field in the high-resolution images of the area obtained with the JunoCam instrument onboard the Juno spacecraft about 28 hours after the impact \citep{Arimatsu2022-if}.

The diameter of the PONCOTS flash object ($D = 23.5^{+3.1}_{-1.0}~{\rm m}$) is approximately 2 times larger than that of the second largest impact (the 2016 March impact, $D = 10 \-- 13~{\rm m}$) and 2.6 times larger than the average diameter of the other six impacts ($D \sim 9.0 ~{\rm m}$).
It is interesting to note that the ratio of the number of impact objects $N$ with $D \gtrsim 23.5~{\rm m}$ to those with $D \gtrsim 9.0~{\rm m}$ class impacts ($1/7 \sim 0.14$) is consistent with that expected from the size distribution typically assumed for Jupiter family comets impacting Jupiter
($N(\gtrsim 23.5~{\rm m})/N(\gtrsim 9.0~{\rm m}) = (23.5/9.0)^{-2} \sim 0.15$; e.g., \citealt{Levison2000-ze}).

The details of each impact event are presented as follows.

% --------------------------------------------------------------------------------------------------------------------------------
\begin{table*}[]
\caption{(Re-)analysis results of the Jovian impact flashes using our cloud reflection model}
\centering 
\begin{tabular}{l c l  cc cc cc cc}
\hline\hline
Date &  &  & \multicolumn{2}{c}{luminous energy} & \multicolumn{2}{c}{kinetic energy} & \multicolumn{2}{c}{mass}  & \multicolumn{2}{c}{diameter(*5)}  \\
(yr-mm-dd) & $\mu$  & $f_{\rm CR}$  & \multicolumn{2}{c}{$E_0$  ($10^{14}$ J)}  & \multicolumn{2}{c}{$E_T$ ($10^{14}$ J)} & \multicolumn{2}{c}{$M$ ($10^5$ kg)}  & \multicolumn{2}{c}{$D$ (m)}             \\
      &  &  &  original & revised &  original & revised  &  original & revised &  original & revised \\
\hline
2010-06-03(*1)  &  0.62 &  1.2 (B) & $0.3 \-- 2.7$ & $0.2 \-- 1.7$ & $1.9 \-- 14$ & $1.1 \-- 8.9$ & $1.1 \-- 7.8$ & $ 0.6 \-- 5.0$  & $7.0 \-- 14$ & $5.8 \-- 12$  \\ 
             &      &  1.4 (R)&               &                 &             &                    &         &  \\
            &       &       &                &                 &         &        &         &  \\             
2010-08-20(*1)  &  0.90 &  1.7 & $0.6 \-- 2.0$  & $0.3 \-- 1.0$  & $3.7 \-- 11$  &  $1.9 \-- 5.6$ & $2.1 \-- 6.1$ & $1.1 \-- 3.1$  &  $8.7 \-- 13$  & $7.0 \-- 10$  \\
2012-09-10(*1)  &  0.51 &  1.2 & $1.6 \-- 3.2$  & $0.9 \-- 1.9$ &  $9.0 \-- 17$  &  $5.4 \-- 10$  & $5.0 \-- 9.5$ & $3.0 \-- 5.6$  &  $12 \-- 14$  &  $9.9 \-- 12$ \\
2016-03-17(*2)  &  0.05 &  0.2 & $1.3 \-- 2.8$  & $1.1 \-- 2.4$  &  $7.3 \-- 14$ &  $6.3 \--13$   & $4.0 \-- 8.1$ & $3.5 \-- 6.9$  &  $11 \--14$   &  $10 \--13$\\  
2017-05-26(*2)  &  0.60 &  1.2 & $0.2 \-- 0.4$ & $0.1 \-- 0.2$  &  $1.3 \-- 2.3$  & $0.8 \-- 1.5$ & $0.8 \-- 1.3$ & $0.4 \-- 0.8$  & $6.1 \-- 7.4$ &  $5.3 \-- 6.4$ \\  
            &       &       &     &       &    &   \\
2019-08-07(*3)  &  0.57 &  1.4 & $0.7 \-- 1.1$  & $0.4 \-- 0.6$ & $4.0 \-- 6.3$  & $2.4 \-- 3.8$ & $2.2 \-- 3.5$ & $1.3 \-- 2.1$ & $8.9 \-- 10$ & $7.5 \-- 8.7$ \\
             &      &      &      &   &    &    \\
2021-10-15(*4)  &  0.91 &  1.9  (V)  & $\--$ &   $18^{+9}_{-2} $ & $\--$ & $74^{+33}_{-9}$ & $\--$  & $41^{+19}_{-5} $  & $\--$ & $23.5^{+3.1}_{-1.0}$\\
            &       &  1.7  (Gh)     &        &    &    &  &    & \\
            &       &  0.4 (${\rm CH_4}$)     &     &       &    &  &    & \\
\hline
\end{tabular}
\begin{flushleft}
Note: (*1) Original values from \cite{Hueso2013-ki}. 
(*2) Original values from \cite{Hueso2018-nr}.
(*3) Original values from \cite{Sankar2020-lm}. 
%(*4) Values from \cite{Arimatsu2022-if}.
(*4) Values from \cite{Arimatsu2022-if}, which already incorporated this cloud correction factor following the methodology presented in this study.
(*5) Mass density $\rho$ is assumed to be $\rho = 600~{\rm kg \, m^{-3}}$.
\end{flushleft}
\label{tab:Analysis}
\end{table*}
% --------------------------------------------------------------------------------------------------------------------------------

\subsubsection{2010 June flash}
On June 3, 2010, the first impact flash event on Jupiter was discovered by two amateur astronomers, as reported by \citet{Hueso2010-xv}.
They observed the same flash at different wavelength bands with an effective wavelength of $\lambda = 435$ nm (hereafter "B-band") and $650$ nm ("R-band").
The previous $E_0$ and $E_T$ values were estimated by \citet{Hueso2013-ki} to be $E_0 = 0.3 \-- 2.7 \times 10^{14}~{\rm J}$ and $E_T = 1.9 \-- 14 \times 10^{14} ~{\rm J}$, respectively,
assuming the cloud correction factor to be $f_{\rm CR} = 0.3$ and a blackbody flash radiation with a temperature range of $3500 \-- 10000$ K.
Using the emission angle for this event ($\mu = 0.62$) and the system responses for the two observations provided by \citet{Hueso2013-ki},  
$f_{\rm CR}$ for the B- and R-bands are derived to be $f_{\rm CR} \sim 1.2$ and $f_{\rm CR} \sim 1.4$, respectively, 
which correspond to $F_{\rm cor} = 0.45$ and $0.42$, approximately $40\--50\%$ smaller than the previously assumed $F_{\rm cor} \sim 1/1.3$.
Since $f_{\rm CR}$ for each wavelength band depends on the spectrum of the flash, it varies with the assumed effective temperature. 
However, the variation of $f_{\rm CR}$ is less than $10\%$ in the assumed temperature range ($3500 \-- 10000$ K).
With the $f_{\rm CR}$, we re-scale $E_0$ for each band and obtain its upper and lower limits.
We then derive the revised $E_T$, $M$, and $D$ using the relationships presented in Sect.~\ref{sub:PONCOTS}.
The obtained upper- and lower-limit values for the bulk parameters are $E_0 = 2 \-- 17 \times 10^{13} \, {\rm J}$, $E_T = 1.1 \-- 8.9 \times 10^{14} \, {\rm J}$, and $D = 5.8 \-- 12 ~ {\rm m}$, 
which are approximately $70\%$, $60\%$, and $20\%$ smaller than the previous estimate, respectively.

\subsubsection{2010 August -- 2017 May flashes}
\citet{Hueso2013-ki,Hueso2018-nr} reported five impact flash events detected between 2010 and 2017, four of which are newly reported.
From the five reported detections and their estimated sizes of the impact objects, 
they performed order-of-magnitude estimates of the impact rates of meter to decameter-sized outer solar system bodies on the Jovian surface.
In these studies, they use a constant $f_{\rm CR} = 0.3$, except for the March 2016 event, which occurs close to the Jovian limb and the contribution of the cloud reflection was not expected.

We estimate $f_{\rm CR}$ for each detection case using our cloud reflection model and derive the revised bulk parameters, which are presented in Table~\ref{tab:Analysis}.
The assumed temperature ranges for the deviations of $E_0$, and $E_T$ are $3500 \-- 10000 ~ {\rm K}$ for the 2010 August and 2012 September cases and $3500 \-- 8500 ~ {\rm K}$ for the 2016 March and 2017 May cases,
respectively, following \citet{Hueso2013-ki,Hueso2018-nr}.
As in the case of the 2010 June flash, the revised $E_0$ and $E_T$ for these flashes are up to $\sim 50\%$ smaller than the previous estimates due to underestimations of the cloud reflection contributions.
On the other hand, the revised $D$ values are up to $\sim 20\%$ smaller than the original ones and therefore do not significantly change the previous results of the impact size frequency studies. 

\subsubsection{2019 August flash}
Another impact flash event on 7 August 2019 was reported by \citet{Sankar2020-lm}.
Since the single wavelength movie data of this event were recorded with better image quality than the previous ones, the luminous and kinetic energies were obtained with small uncertainties under an assumed effective temperature range of $3500 \-- 10000 ~ {\rm K}$.
With the contribution factor $f_{\rm CR} \sim  1.4$, $E_0$ and $E_T$ are corrected to be $0.4\-- 0.6 \times 10^{14}$ and $2.4 \-- 3.8 \times 10^{14}$ J, respectively, 
which is an approximately factor of two smaller than the original values.
In contrast to the detection cases presented above, 
these corrected energy ranges are outside the uncertainty ranges of the original values. 
%They may partially modify the ablation and fragmentation model results and discussions provided by \citet{Sankar2020-lm}.

\section{Conclusions} 
\label{sec:concl}
The cloud reflection component can account for up to $70\%$ of the observed flash brightness, and its fraction varies strongly with the observation geometry. 
The application of our proposed correction method to the previously reported flashes provided the revised bulk parameters of the impact objects.
These revised parameters did not significantly affect the discussion of the impact object sizes raised in the previous studies.
However, the proposed cloud reflection correction will be critical for ongoing more detailed investigations, 
including present-day ablation and fragmentation modelling, 
of the flashes observed with current high-sensitivity and multi-wavelength systems such as PONCOTS.
Also, the application of the present correction method to the ultraviolet (UV) spectrum of the impact flash recently obtained with the Ultraviolet Spectrograph instrument onboard the Juno spacecraft \citep{Giles2021-sj} may influence discussions of its emission characteristics.
However, the reflection component in the UV wavelength range is expected to be much weaker than that in the visible range due to the low UV reflectivity of the Jovian surface clouds (e.g., \citealt{Giles2021-he}). 
Further discussion will require additional detailed modelling of the reflection component, taking into account the UV reflective properties.

It should be noted that the exact evaluation of the reflection component is a more complex problem of three-dimensional radiative transfer and requires consideration of spatial variations of the optical properties of Jovian clouds.
Although the present model is still a kind of simple approximation and therefore imperfect, it provides a practical method for deriving the most appropriate contribution factors with limited observational and computational resources. 
Most of the recent Jovian flashes have been observed with Bayer filter colour cameras or multiband imaging instruments \citep{Hueso2018-nr,Arimatsu2022-if}. 
The spatially resolved surface colour information obtained simultaneously with these instruments would be a key to understanding a more detailed reflection characteristic of the impact site.

In addition to the impact flash studies, 
our reflection model will be useful for understanding other optical transients of the Jovian upper atmosphere, 
such as the sprites-like events recently discovered by \citet{Giles2020-rr}.
Non-negligible contributions from the cloud reflection components must be present in such observed transients.
The present model is therefore expected to have a broader application to optical transients occurring on the Jovian surface.  

\begin{acknowledgements}
We thank Ricardo Hueso for impartial review and for providing constructive suggestions. 
We thank Erich Karkoschka for providing the data on the methane absorption model, and Liming Li for providing the data on the phase-angle dependence of Jovian albedo. 
This research has been partly supported by JSPS grants (18K13606, 21H01153).
\end{acknowledgements}

\bibliographystyle{aa} % style aa.bst
\bibliography{ref01} % your references Yourfile.bib
\end{document}